# Interest of Integrating Spaceborne LiDAR Data to Improve the Estimation of Biomass in High Biomass Forested Areas

Mohammad El Hajj [1,*], Nicolas Baghdadi [1], Ibrahim Fayad [1], Ghislain Vieilledent [2,3], Jean-Stéphane Bailly [4] and Dinh Ho Tong Minh [1]

[1] Institut national de recherche en sciences et technologies pour l'environnement et l'agriculture (Irstea), Unité Mixte de Recherche (UMR) Territoires, Environnement, Télédétection et Information Spatiale (TETIS), 500 rue Jean François Breton, 34093 Montpellier CEDEX 5, France; nicolas.baghdadi@teledetection.fr (N.B.); ibrahim.fayad@teledetection.fr (I.F.); dinh.ho-tong-minh@irstea.fr (D.H.T.M)

[2] Centre de coopération internationale en recherche agronomique pour le développement (Cirad), Unité Propre de Recherche (UPR) Forêts et Sociétés (F&S), F-34398, Montpellier, France; ghislain.vieilledent@cirad.fr

[3] Joint Research Center of the European Commission, Bio-economy unit, I-21027 Ispra, Italy

[4] AgroParisTech, Unité Mixte de Recherche (UMR) Laboratoire d'étude des interactions Sol-Agrosystème-Hydrosystème (LISAH), 2 place Pierre Viala, 34060 Montpellier, France; bailly@agroparistech.fr

* Correspondence: mohammad.el-hajj@teledetection.fr; Tel.: +33-467-548-724

**Abstract:** Mapping forest AGB (Above Ground Biomass) is of crucial importance to estimate the carbon emissions associated with tropical deforestation. This study proposes a method to overcome the saturation at high AGB values of existing AGB map (Vieilledent's AGB map) by using a map of correction factors generated from GLAS (Geoscience Laser Altimeter System) spaceborne LiDAR data. The Vieilledent's AGB map of Madagascar was established using optical images, with parameters calculated from the SRTM Digital Elevation Model, climatic variables, and field inventories. In the present study, first, GLAS LiDAR data were used to obtain a spatially distributed (GLAS footprints geolocation) estimation of AGB (GLAS AGB) covering Madagascar forested areas, with a density of 0.52 footprint/km$^2$. Second, the difference between the AGB from the Vieilledent's AGB map and GLAS AGB at each GLAS footprint location was calculated, and additional spatially distributed correction factors were obtained. Third, an ordinary kriging interpolation was thus performed by taking into account the spatial structure of these additional correction factors to provide a continuous correction factor map. Finally, the existing and the correction factor maps were summed to improve the Vieilledent's AGB map. The results showed that the integration of GLAS data improves the precision of Vieilledent's AGB map by approximately 7 t/ha. By integrating GLAS data, the RMSE on AGB estimates decreases from 81 t/ha ($R^2$ = 0.62) to 74.1 t/ha ($R^2$ = 0.71). Most importantly, we showed that this approach using LiDAR data avoids underestimating high biomass values (new maximum AGB of 650 t/ha compared to 550 t/ha with the first approach).

**Keywords:** aboveground biomass mapping; LiDAR; ICESat GALS; field inventories

## 1. Introduction

Monitoring the carbon cycle and carbon stocks is of high importance to understand climate change. Several studies have reported that more than 40% of the world's vegetation carbon stocks is stored in tropical forests [1,2]. In tropical forests, the quantity of carbon represents 43% to 55% of Above Ground Biomass (AGB) [3–5]. Thus, mapping the AGB of tropical forests is of great importance in monitoring carbon stocks. Field inventories for AGB estimates, either by destructive (cutting and then weighing the tree) or non-destructive methods (by means of allometric equations), provide good estimates. However, these methods are not operational because they involve a great deal of labor and time and allow AGB estimates only at a local scale. Thus, a forest cannot be mapped using field inventories, hence the importance of remote sensing technology that facilitates the mapping of AGB. Indeed, remote sensing technology provides data for AGB estimates that cover large areas with a high spatial resolution and high revisit time.

Three main remote sensing data types are used for AGB estimates: optical, SAR (Synthetic Aperture Radar), and LiDAR. Optical images at low or medium resolutions and radar backscattering coefficient data are robust enough to estimate low to medium level AGB due to saturation of remote sensing data. Zhao et al. [6] and Lu et al. [7] have shown that optical data allow AGB estimates until AGB levels between 55 and 159 t/ha, depending on the forest species composition. In addition, SAR amplitude data, mainly in the L-band, were used to estimate the AGB. Luckman et al. [8] observed a saturation point of 60 t/ha when plotting the JERS L-band backscattering coefficients as a function of the forest biomass located in the Central Amazon Basin. Baghdadi et al. [9] found that the ALOS/PALSAR L-band backscattering coefficients saturate when the biomass of the Brazilian eucalyptus plantations reaches 50 t/ha. The use of radar backscattering coefficients in the P-band allows the estimation of higher AGB levels (290 t/ha for P-band [10]). However, to date, there are no available P-band SAR instruments operating from space, and the airborne P-band SAR data are commercial, which makes the use of these sensors expensive. The near future space-borne P-band SAR sensor (BIOMASS mission scheduled to launch in 2020) would allow tomographic analyses of SAR data for higher level AGB estimates [11]. Since Reigber and Moreira [12], the exploitation of SAR data for conducting tomographic analyses has been the object of a growing interest within the SAR community. By using tomography, forest biomass can be investigated by considering not only the backscatter at each slant range and azimuth location, but also its vertical distribution. The potential of tomography to characterize forest structure was previously assessed in a number of studies relating the vertical structure of forests to forest AGB over French Guiana [13–15]. In these studies conducted in French Guiana, the SAR signal in the P-band coming from upper vegetation layers (determined using SAR tomographic analyses) was found to be strongly correlated with forest AGB for AGB values ranging from 200 t/ha to 500 t/ha [13,15]. This finding was the first demonstration that forest AGB can be determined up to 500 t/ha with a 10% error at the 4-ha scale [14].

Currently, LiDAR is the only available technology able to estimate higher AGB levels (up to 1200 t/ha from airborne LiDAR) [16,17] in comparison to optical and SAR amplitude data. LiDAR data capture the vertical structure of trees and allow the estimation of tree height up to 40 m with good precision [18–20]. The tree height derived from LiDAR is strongly correlated with the AGB of the trees, with no saturation at higher AGB values [19,21,22]. LiDAR data can be acquired from an aircraft and from space. Airborne and spaceborne LiDAR sensors record waveforms from small (<1 m) and large footprints (up to 60 m), respectively. Several studies have shown that the estimation of AGB from airborne LiDAR data is more accurate than that from spaceborne LiDAR [23,24]. However, the acquisition of airborne LiDAR data is costly, and the spatial coverage is limited to small areas. On the other hand, the available space-borne LiDAR data acquired by the Ice Cloud and Land Elevation Satellite (ICESat) are free, but do not provide continuous coverage of the earth. To overcome the limitation of spatial cover of LiDAR data and the saturation of optical and SAR amplitude data at medium AGB values, several studies tend to combined LiDAR with optical or SAR data for continuous AGB mapping at regional and global scales.

At the regional scale, Mitchard et al. [21] estimated the AGB in Lopé National Park in central Gabon by coupling GLAS, PALSAR (L-band), and SRTM data. Lorey's height was first derived from GLAS data and then converted to AGB through a simple equation. This equation was fitted using plot field measurements of Lorey's height and AGB. Furthermore, a classification (40 classes) was performed using radar and SRTM data to determine regions with homogeneous vegetation. Finally, GLAS AGB estimates located within each region were averaged, enabling the spatial extrapolation of AGB estimates. The results showed relative error of AGB estimates of ±25% (AGB between 50 and 900 t/ha). Asner et al. [25] mapped the Aboveground Carbon Density "ACD" (ACD = 0.47 × AGB) in one northern (659,592 ha) and one southern (1,713,088 ha) region of Madagascar using airborne LiDAR data, SRTM derived variables, and optical images. First, ground-based ACD estimate plots located within all forest types were used to calibrate LiDAR data to the ACD. Later, the airborne LiDAR-derived ACD was related to the SRTM-derived variables and variables derived from optical data through a linear regression model. Finally, this linear model was applied to map the ACD at 1 ha resolution in the two regions. The results showed that the uncertainty of AGB estimates is equal to 35% and 10% in the northern and southern regions, respectively (ACD between approximately 5 t/ha and 300 t/ha in both regions).

At the global scale, Saatchi et al. [22] mapped the AGB of world tropical forests using a combination of data from 4,079 in situ inventory plots (across the three tropical continents) and GLAS samples of forest structure, plus optical and microwave imagery with 1-km spatial resolution. In this study, a power-law functional relationship ($R^2$ = 0.85) between the in situ Lorey's height and in situ AGB was first performed. This relationship was then applied to tree height derived from GLAS to estimate AGB at each GLAS footprint location. Finally, a fusion model based on the maximum entropy (MaxEnt) approach was performed using spatial imagery to extrapolate AGB measurements from inventory plots and GLAS footprints to the entire landscape. Baccini et al. [26] derived a carbon density map of pan-tropical forests using GLAS data together with MODIS images (Bidirectional Reflectance Distribution Function and land surface temperature) and SRTM data. In this this study, in situ AGB was first derived from plots within GLAS footprints using trees characteristics. Then, a statistical relationship between the in situ AGB estimates and GLAS waveform metrics was established, allowing the estimation of AGB for all GLAS footprints located across the tropics. Finally, a model relating GLAS-based AGB estimates and MODIS and SRTM data was calibrated and applied to derive the AGB map. Mitchard et al. [27] assessed the reliability of a global AGB map produced by Saatchi et al. [22] and Baccini et al. [26] by using an accurate AGB map of Amazonian Columbia as a reference dataset. Mitchard et al. [27] observed substantial discrepancies between the maps of Saatchi et al. [22] and Baccini et al. [26] over tropical forest areas (up to ±150 t/ha), even though both maps give similar means and total AGB values on the continent scale. In addition, the maps of Saatchi et al. [22] and Baccini et al. [26] have higher AGB values (up to 150 t/ha) in comparison to the accurate AGB map of Amazonian Columbia. Such bias could be related to the saturation of spatial data and to the use of an insufficient number of high in situ AGB values during model calibration, which reduces model performance for the estimation of high AGB values.

The main goal of this study is to investigate the contribution of spaceborne LiDAR data in overcoming the saturation at high AGB values of existing map produced in Madagascar by Vieilledent et al. [28] using optical satellite images, a Digital Elevation Model (DEM) and climatic variables. To produce its AGB map, Vieilledent et al. [28] first use a random forest model to relate in situ AGB measurements to the EVI (Enhanced Vegetation Index) derived from optical images, parameters calculated from the SRTM Digital Elevation Model, and climatic variables. Then, this random forest model was applied to map the AGB of forested areas in Madagascar. To date, the Vieilledent AGB map is the most recent and accurate map with medium resolution (250 m) for forested areas in Madagascar. The inconvenience of the Vieilledent AGB map is the inability to measure high AGB values, and therefore, a new method is required that incorporates LiDAR remote sensing to overcome such inconvenience. This inconvenience is mainly due to the use of the EVI and the percent tree cover derived from optical images at medium resolution as predictive variables for AGB estimates. The optical data at medium resolution saturates at high AGB values and induces an underestimation of high AGB [6,7]. An improvement in Vieilledent's AGB map will allow a more accurate estimation of carbon stocks and mapping in forested areas in Madagascar. In our study, The AGB was first estimated from GLAS and DEM metrics, providing a spatially distributed (GLAS footprints geolocation) AGB estimation (GLAS AGB). Second, the spatial dependency of the additional correction factors (Vieilledent's AGB map—GLAS AGB at each GLAS footprint location) was modeled, and an ordinary kriging interpolation of additional correction factors was performed to provide a correction factor map. Finally, the correction factor map and Vieilledent's AGB map were summed to improve Vieilledent's AGB map, taking into account the addition of GLAS data in AGB estimation. A description of the study area and the different datasets used in this study is provided in Section 2. Section 3 presents the methodology. The results and discussions are shown in Sections 4 and 5, respectively. Finally, Section 6 presents the conclusion.

**2. Study Area and Datasets**

*2.1. Study Area*

Madagascar is an island country in the Indian Ocean located to the southeast of the African continent (latitude between 12°S and 26°S, Figure 1). The total area of the country is approximately 58,154,000 ha, out of which 12,553,000 ha are forests, according to FAO estimates in 2010 (U.N. FAO). Between 1990 and 2010, Madagascar lost 8.3% of its forest cover (approximately 1,139,000 ha) [1]. Madagascar is composed of three ecoregions, with three forest types defined according to climate and vegetation type [29]. The eastern, western and southern parts contain mainly moist, dry and spiny forests, respectively [29] (Figure 1). The climates of the eastern, western and southern ecoregions are humid, temperate, and arid, respectively. This difference in climate is caused mainly by the Indian Ocean trade winds, which bring with them variations in precipitation throughout the region. The terrain in Madagascar is sloping. The slope of the eastern ecoregion reaches 25% and rises occasionally to reach 30%. The western and southern ecoregions are less sloping, with global slopes of less than 10%.

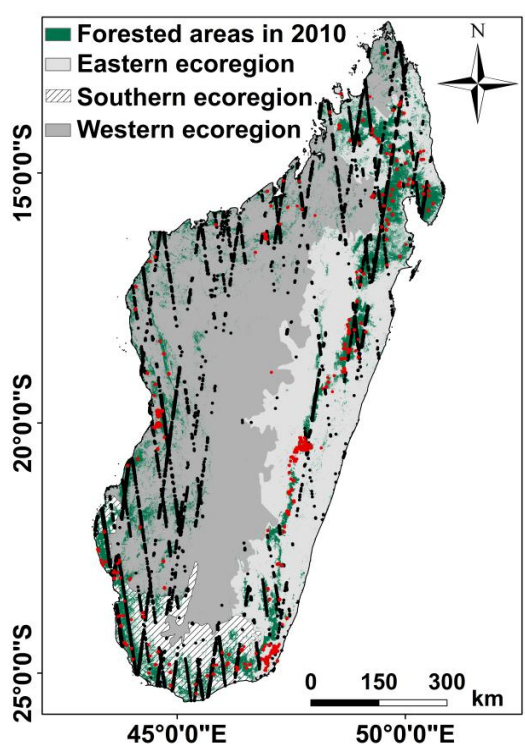

**Figure 1.** Madagascar, with the three ecoregions; forested areas in 2010 determined from Vieilledent's Above Ground Biomass (AGB) map. Black points are the 48247 Geoscience Laser Altimeter System (GLAS) footprints located within forested areas. Red points correspond to field inventories.

*2.2. Datasets*

2.2.1. In Situ AGB Data

Nine forest inventories have been conducted to measure the AGB between 1995 and 2013 (1995, 1996, 2007, and 2009 through 2013) in 1771 field plots. The dimensions of the plots were 0.13 ha (radius = 30 m) in the moist forest, and 0.28 ha (radius = 20 m) in the dry and spiny forests. The AGB was computed for each tree (*i*) using the allometric equation defined by Chave et al. [30]:

$$AGB_i = 0.0673 \times (\varrho_i\, D_i\, H_i)^{0.976} \tag{1}$$

where $\varrho_i$ is the tree wood density (g·cm$^{-3}$), $D_i$ is the tree diameter (cm) at a height of 130 cm, and $H_i$ is the tree height (m). Detailed descriptions about in situ AGB measurement procedures are given in the studies of Vieilledent et al. [28,31].

The field inventories used in this study are those utilized by Vieilledent et al. [28], except those that were disturbed between 1995 and 2013, which were eliminated through an analysis of forest maps in 2000 derived by Harper et al. [32] and by the photo-interpretation of Landsat images time series between 2000 and 2013. The number of field inventories that were not disturbed between 1995 and 2013 is 1194 (Figures 1 and 2).

Field inventory methods for AGB estimates allow precise AGB estimates only at the local scale. Thus, mapping the AGB and calculating the total AGB for given wide forested areas could not be performed only by means of field inventories. Spaceborne LiDAR have the potential to map AGB, since these sensors provide precise spatially distributed information about tree height, which is well correlated to AGB.

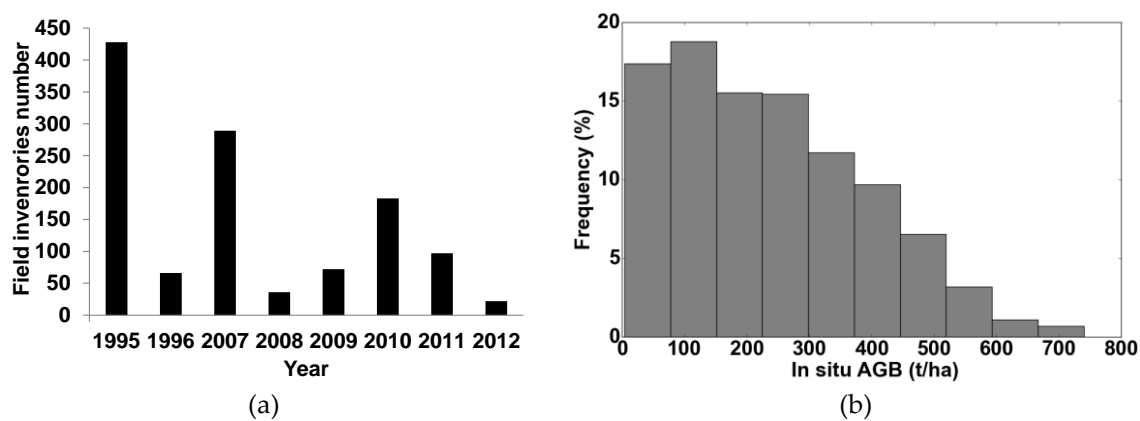

(a)      (b)

**Figure 2.** (**a**) Number of field inventories for a given year; (**b**) distribution of AGB density from field inventories.

2.2.2. Vieilledent's Aboveground Biomass Map

Vieilledent's AGB map [28] provides the AGB for forests in Madagascar in 2010, with a spatial resolution of 250 m. Vieilledent's AGB map was derived from three types of explicative variables (vegetation indices, topography, and climatic data) using the Random Forest (RF) regression technique. Vegetation indices are the EVI (Enhanced Vegetation Index) and the percent tree cover (%VCF), both derived from MODIS (Moderate Resolution Imaging Spectroradiometer) satellite remote sensing imagery acquired between 2000 and 2010. A topographic variable (elevation) was obtained from the 30-m resolution SRTM (Shuttle Radar Topography Mission) global elevation dataset. Climatic data (1950–2000) are the cumulative annual precipitation, mean annual temperature, and temperature seasonality (standard deviation of monthly temperature × 100), all of which are extracted from the MadaClim website (http://madaclim.cirad.fr).

To derive Vieilledent's AGB map, in situ AGB were first related to explicative variables by means of the RF regression technique. Then, the resulting model was applied to map AGB in forested areas in Madagascar with spatial resolution of 250 m. Climatic variables were the most important, compared with EVI, %VCF, and elevation. In the study of Vieilledent et al. [28], a cross-validation procedure was used to validate the AGB map using field inventories as the reference dataset.

2.2.3. LiDAR Data

LiDAR data were acquired between 2003 and 2009 by the Geoscience Laser Altimeter System (GLAS) sensor on board the Ice, Cloud, and Land Elevation (ICESat) Satellite. The GLAS sensor operates in the near-infrared (1064 nm) wavelength and illuminates footprints with a nearly circular shape that are approximately 70 m in diameter. GLAS LiDAR data are free and available for all continents. Footprints are separated by approximately 172 m in the along-track direction. The horizontal geolocation accuracy of the GLAS footprints is 3.7 m (on average), and the vertical accuracy is between 0 and 3.2 cm over flat surfaces, on average [33,34]. Only the GLA01 and GLA14 data products available from ICESAT/GLAS were used in this study. The GLA01 product contains the full recorded waveform data. The GLA14 product, derived from the GLA01 product, contains several useful data for each footprint, such as the cloud flag index, saturation waveform index, land surface elevation from SRTM, centroid elevation derived from the waveform, and background noise.

To eliminate unreliable GLAS data (i.e., data affected by atmospheric conditions), several filters were applied: (1) footprints with associated centroid elevations significantly different than the corresponding SRTM elevations were excluded (|GLAS − SRTM| > 100 m); (2) footprints corresponding to waveforms with a low signal to noise ratio (SNR) were also removed (SNR < 15) [33]; (3) saturated waveforms were eliminated (saturation index satNdx # 0); and (4) only the cloud-free footprints were conserved (cloud flag FRir_qaFlag = 15). In addition, GLAS footprints located inside forest stands (selected using the existing AGB map) were conserved. From the original database of 1,772,000 footprints, 48,247 footprints that respect all criteria mentioned above were kept (Figure 1). The density of GLAS footprints in forested areas is 0.52 points/km$^2$.

Moreover, metrics were derived from reliable GLAS waveforms provided in the GLA01 product to represent the vertical variables of the canopy. These metrics are the Waveform extent (Wext), percentile heights (H) of GLAS waveforms (10 through 90%, with steps of 10%), Leading Edge (LE), and Trailing Edge (TE). A noise threshold equal to 4.5 times the standard deviation of the background noise was used to determine the waveform beginning and end [35]. The waveform extent is the difference between the signal end and signal start. The waveform extent was corrected for slope effects using the following equation [36,37]:

$$Wext\_cor = Wext \times 0.15 - 0.5 \times d \times \tan(\theta) \qquad (2)$$

where d is the footprint diameter (in m), and θ is the mean slope of the illuminated surface area. Wext is expressed in ns and Wext_cor is in m.

In addition, the Gaussian peaks resulting from the decomposition of each GLAS waveform, which represent canopy features, such as canopy top, canopy trunks, or ground, were identified. In this study, the first Gaussian peak was considered as the top of canopy return, and the stronger of the last two Gaussian peaks was selected as the ground return [38]. After identifying the top and ground peaks, the percentile heights of GLAS waveforms (10 through 90%, with steps of 10%) were also calculated from the signal beginning. Finally, the leading edge, defined as the elevation difference between the signal start and the canopy peak's center, and the trailing edge, defined as the difference between the signal end and the ground peak's center [18], were estimated.

2.2.4. Digital Elevation Model

The SRTM (Shuttle Radar Topography Mission) Digital Elevation Model (DEM) with a spatial resolution of 30 m was used is this study. Three variables were derived from the DEM: slope (θ), Terrain Index (TI), and surface Roughness (Roug). The TI map was obtained by calculating the difference between the highest and lowest altitude in a 3 × 3 pixel moving window. The surface Roug map was obtained by computing the standard deviation of the elevation in a 3 × 3 pixel moving window. All SRTM derived variables (θ, TI, and Roug) were resampled (averaging the value of the cells) to 250 m (Vieilledent's AGB map resolution).

**3. Methodology**

The methodology to improve the precision of Vieilledent's AGB map consisted of (1) finding a model predicting in situ AGB from both GLAS and DEM metrics using in situ AGB neighboring

GLAS footprints at a distance of up to 250 m; (2) applying the previous model to all GLAS footprints to derive the AGB (GLAS AGB); (3) calculating the additional correction factors, which were the differences between Vieilledent's AGB map and GLAS AGB, at each footprint location; (4) performing an ordinary kriging interpolation to map the additional correction factors; and (5) improving Vieilledent's AGB map by adding the kriged additional correction factors to Vieilledent's AGB map. It should be noted that our methodology is not over reliant on the existing AGB map (Vieilledent's AGB map) because the data used to produce this AGB map is available for free with global coverage, and therefore it was possible to reproduce Vieilledent's AGB map. A brief scheme explaining the procedures for the improvement of Vieilledent's AGB map is shown in Figure 3.

To assess the relevance of our approach, first, the improved AGB map was compared to the in situ AGB to determine the gain in precision brought to Vieilledent's AGB map. Then, the accuracy of the improved AGB map was compared to the accuracy of (1) the most recent pan-tropical AGB map produced by Avitabile et al. [39] (Avitabile's AGB map); and (2) another AGB map computed in this present study using our database (48,247 GLAS-derived AGB, DEM metrics, and field inventories) following the method proposed by Baccini et al. [26], called "Baccini's approach AGB map".

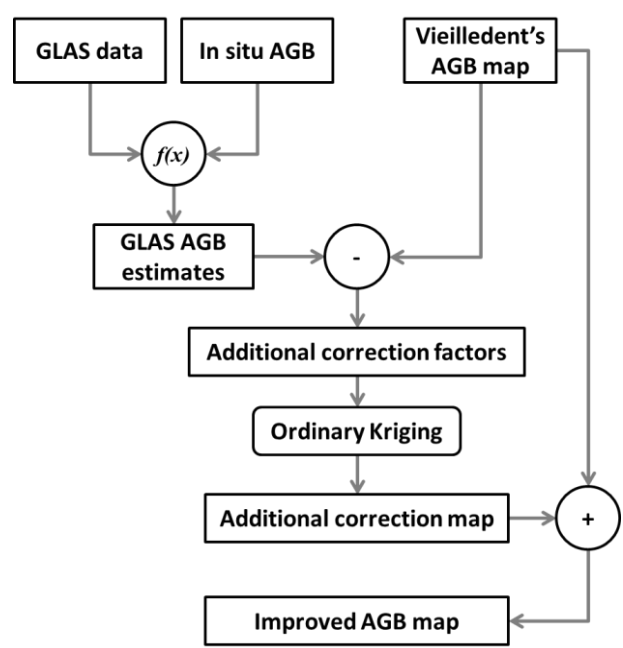

**Figure 3.** Procedure for improving Vieilledent's AGB map.

*3.1. Estimation of the AGB from GLAS Data*

Madagascar is composed of three climatic ecoregions with three different forest types (Figure 1). The western and southern ecoregions have lower in situ AGB values (<150 t/ha on average) compared to the eastern ecoregion (>250 t/ha on average). Accordingly, two different multilinear models were built to relate the in situ AGB to the GLAS metrics (Wext_cor, LE, TE, H10 through H90 with a 10% step) and DEM data (slope, TI, and Roug). The first model links the in situ AGB from the eastern ecoregion to the GLAS and DEM metrics. Similarly, the second model relates the in situ AGB from the western and southern ecoregions to the GLAS and DEM metrics. A step-wise regression technique (with both forward and backward processes) was used to select the best variables to be used for AGB estimation. Finally, these multilinear models were applied to all GLAS footprints, using only best variables to derive the AGB (GLAS AGB). GLAS footprints did not intersect in situ AGB. To associate an in situ AGB value with GLAS footprints, we considered a maximum of 250 m between the in situ AGB and GLAS footprints.

*3.2. The Improved AGB Map*

An ordinary kriging (OK) interpolation was used to improve the precision of Vieilledent's AGB map. The OK model allows the interpolation of additional correction factors (Vieilledent's AGB map—GLAS AGB at each GLAS footprint locations) based solely on a regionalized linear model known as a semivariogram. The semivariogram describes the spatial dependency between additional correction factors and draws the semi-variance $\gamma$ as a function of the distance between samples $h$ using the following function:

$$\gamma(h) = \frac{1}{2N(h)} \sum_{i=1}^{N(h)} [e(s_i) - e(s_i+h)]^2 \qquad (3)$$

where $\gamma(h)$ is the semi-variance as a function of the lag distance $h$, $N(h)$ is the pairs data number separated by h, and e is a local measure of the additional correction factors at locations $s_i$ and $s_i$+h. The semivariogram function has three main parameters: (1) the nugget: the semi-variance value at h close to zero; (2) the sill: semivariance at which no spatial correlation exists at long distances [40]; and (3) the range: the distance at which the sill is reached.

After drawing the empirical semivariogram samples, and assuming an order-2 stationary process (fixed Esperance and homogeneous spatial dependency over all space), an admissible model in $R^2$ is fitted to the empirical variogram, determining the semivariogram function parameters. Ordinary kriging (centered on an unknown value) is thus performed using a fitted semivariogram function, which estimates the value $e$ and the prediction variance at any location s0 (location where no additional correction factors are available) using the linear equation:

$$ẑê(s_0) = \sum_{i=1}^{n} \lambda_i e(s_i) \qquad (4)$$

where $ê(s_0)$ is the predicted value at an unvisited location $s_0$ and $\lambda_i$ are the kriging weights of n neighboring samples [40]. The weights $\lambda_i$ depend on the fitted semivariogram function, the distance to the predicted location, and the spatial design of $e$ data.

From that framework, a sub-variogram model was built for each ecoregion to respect the conditions of stationarity. Then, for each ecoregion, an OK interpolation was performed using the additional correction factors within that ecoregion to create a correction factor map. Furthermore, the correction factor map of that ecoregion was adding to the part of Vieilledent's AGB map that overlaps to increase its precision. Finally, the two improved AGB maps (of both eastern and western ecoregions) derived using the two variograms were combined to obtain the improved AGB map that covered all of Madagascar with a spatial resolution of 250 m.

To calculate the precision of the improved AGB map, first, a spatial intersection between the in situ AGB and the improved AGB map pixels was performed. Then, AGB pixels that contained at least two in situ AGB were selected, as well as the associated in situ AGB. Finally, the RMSE and $R^2$ were calculated by using the averaged values of the in situ AGB located within the same AGB pixel. The in situ AGB used to build the model for AGB estimation from GLAS data were not used to calculate the precision of the improved AGB map.

*3.3. Comparison between the Improved AGB Map and Avitabile's AGB Map*

The precision of the improved AGB map was compared to that of the pan-tropical AGB map produced by Avitabile et al. [39] (Avitabile's AGB map). Avitabile's AGB map was used as benchmark because, to date, this map is considered to be the most recent and accurate pan-tropical AGB map. Avitabile et al. [39] combined the global AGB map of Saatchi et al. [22] and Baccini et al. [26] into a pan-tropical AGB map (1 km resolution) using reference AGB data. The fusion model consists of bias removal and weighted linear averaging of both the Saatchi et al. [22] and Baccini et al. [26] AGB maps to produce an AGB map with higher accuracy. The bias removal consisted of adding the mean difference between the input map and the reference AGB data to the input maps (Saatchi's and Baccini's AGB maps). The results showed that the RMSE of Avitabile's AGB map (89 t/ha) is lower by 15%–21% than that of the input maps (Saatchi's and Baccini's AGB maps).

Avitabile's AGB map is produced with a spatial resolution of 0.00833° (1 km) and the WGS-84 geographic projection. To make Avitabile's and the improved AGB map comparable, Avitabile's AGB map was re-projected into UTM (Universal Transverse Mercator), to be in the same projection as the improved AGB map. In addition, the improved AGB map was resampled to 1 km as follows: the AGB pixels of the improved AGB map that fall within each cell of Avitabile's AGB map were averaged.

Finally, Avitabile's AGB map and the improved AGB map resampled to a spatial resolution of 1 km were compared to the in situ AGB in the same manner adopted to validate the improved AGB map. However, to validate these maps, we used AGB pixels that cover more than three in situ AGB values.

*3.4. Comparison between the Improved AGB Map and the AGB Map from Baccini's Approach*

To assess the relevance of our approach it was important to compare it to Baccini's approach, since the latter is the most commonly used approach for AGB mapping [22,25,41]. First, we used our database (48,247 GLAS derived AGB, DEM metrics, and field inventories) and the auxiliary variables from the study of Vieilledent et al. [28], to produce a AGB map following the method proposed by Baccini et al. [26]. Then, we compared between precision of the improved AGB map and the precision of the AGB map from Baccini's approach. The precision of the AGB map from Baccini's approach was calculated in the same manner used to validate the improved AGB map.

To produce the AGB map using Baccini's approach, first, the established relationships between in situ AGB and both GLAS and DEM metrics (cf. Section 3.1) were applied to derive the AGB from each GLAS datapoint (48,247 footprints). Second, the GLAS-derived AGB values were related to the three types of explicative variables in the study of Vieilledent et al. [28] using the random forest model. Finally, the random forest model was applied for AGB mapping with a spatial resolution of 250 m, and the AGB map from Baccini's approach was obtained.

Baccini et al. [26] provided a pan-tropical AGB map covering Madagascar. In their study, the in situ AGB measurements used to build the relationship with GLAS and DEM metrics are missing over Madagascar. Therefore, Baccini's AGB relationship is not representative enough to derive AGB estimates from GLAS metrics, leading to an AGB map with poor precision. For this reason, Baccini's AGB map was reproduced in the present study using our in situ AGB measurements available for Madagascar.

**4. Results**

*4.1. Estimation of the AGB from GLAS*

14 plots neighboring GLAS footprints at a distance of 250 m in the eastern ecoregion were used to relate the in situ AGB to all GLAS and DEM metrics (Wext_cor, LE, TE, H10 through H90, slope, TI, and Roug). Similarly, 13 plots were used to link the in situ AGB to all GLAS and DEM metrics in the western and southern ecoregions. Considering a distance lower than 250 m to associate an in situ AGB value with GLAS footprints yields an insufficient sample number to build a model that predicts in situ AGB from both GLAS and DEM metrics. As an example, at a distance lower than 100 m, only 6 samples are available.

For the eastern ecoregion, the results showed that the RMSE of AGB estimated using all GLAS and DEM metrics was equal to 56.2 t/ha (Relative Root Mean Square Error "RRMSE" ~20%, $R^2$ = 0.90). For the western and southern ecoregions, the RMSE of AGB estimated from all GLAS and DEM metrics was equal to 20.7 t/ha (RRMSE~34%, $R^2$ = 0.59). Moreover, the results show that for all ecoregions, the most significant variables for AGB estimates, determined using a stepwise regression, were Wext_cor, LE, H80 and TI:

$$AGB_{insitu} = Wext\_cor + LE + H80 + TI \qquad (5)$$

Table 1 shows the coefficients and p-values for each significant variable. The use of significant variables to derive the AGB decreases the error of in situ AGB estimates. For the eastern ecoregion, the RMSE decreased from 56.2 to 51.5 t/ha (RRMSE~18%, $R^2$ = 0.91) when only the significant variables were used (Figure 4a). In addition, for the western and southern ecoregions, the use of significant variables decreased the RMSE of AGB estimates from 20.7 to 18.9 t/ha (RRMSE~31%, $R^2$ = 0.66) (Figure 4b). Finally, these linear regression models were applied using only significant variables to derive the AGB from all GLAS footprints.

Table 1. Coefficients and *p*-values of each significant variable.

|  | Eastern Ecoregion | | Western/Southern Ecoregion | |
|---|---|---|---|---|
|  | Coefficient | *p* Value | Coefficient | *p* Value |
| Wext_cor | 35.13 | 0.002 | 4.66 | 0.033 |
| lead | −14.79 | 0.002 | −3.35 | 0.041 |
| H80 | −11.99 | 0.048 | −5.26 | 0.048 |
| TI | 13.72 | 0.000 | 7.00 | 0.013 |

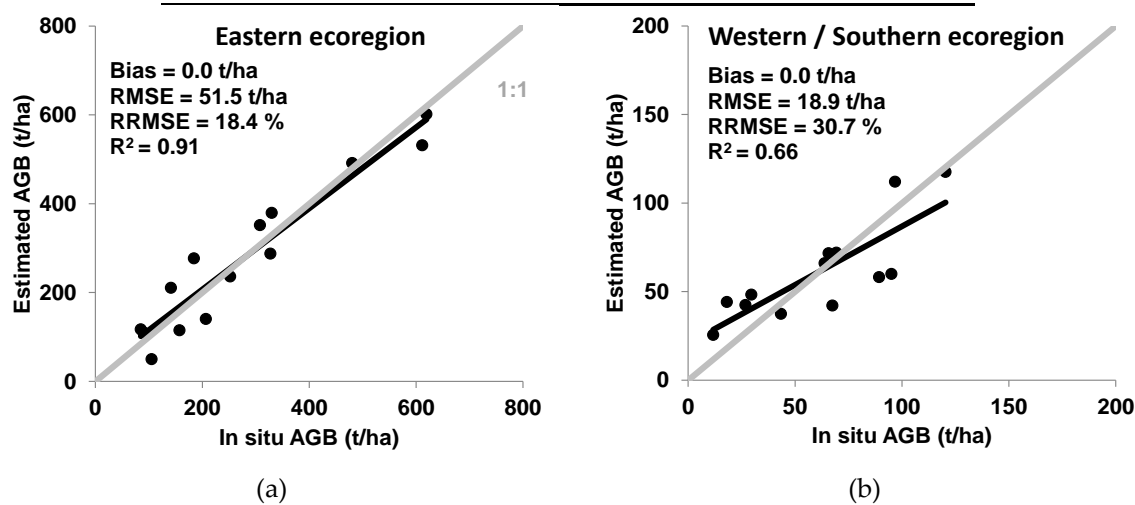

(a)                                      (b)

**Figure 4.** AGB estimation from significant GLAS and Digital Elevation Model (DEM) metrics as a function of in situ AGB in the eastern (**a**) and western and southern ecoregions (**b**).

*4.2. The Improved AGB Map*

To improve Vieilledent's AGB map, the ordinary kriging technique was used. First, the difference (additional correction factors) between the AGB from Vieilledent's AGB map and the GLAS AGB was calculated at the location of each GLAS footprint. Second, the semivariogram of the additional correction factors was performed for each ecoregion separately to respect the conditions of stationarity. Figure 5 shows the semivariogram samples (black points) obtained in the eastern ecoregion (Figure 5a) and in the western and southern ecoregions (Figure 5b). Semivariogram samples were then fitted using an exponential function with a nugget of 17,014 (t/ha)$^2$, partial sill "difference between the nugget and the sill" of 10,710 (t/ha)$^2$ and range of 16,699 m for the eastern ecoregion (Figure 5a), and with a nugget of 197 (t/ha)$^2$, partial sill of 191 (t/ha)$^2$ and range of 261 m for the western and southern ecoregions (Figure 5b). Furthermore, for each ecoregion, the additional correction factors were kriged using the exponential equation with the associated parameters (nugget, partial sill, and range) to provide a correction factor map for that ecoregion. Later, the correction factor map of that ecoregion was added to the corresponding part of Vieilledent's AGB map to increase its precision. Finally, the two sub-improved AGB maps were combined to obtain the improved AGB map covering Madagascar. The improved AGB map has the same resolution as Vieilledent's AGB map (250 m).

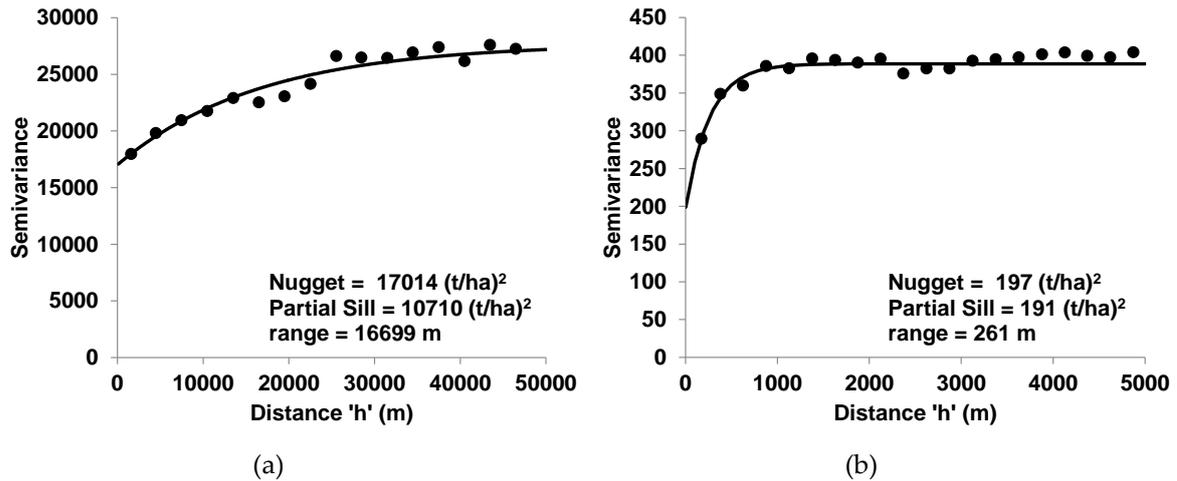

**Figure 5.** Variograms of the additional correction factors with the associated parameters (nugget, partial sill, and range). (**a**) Eastern ecoregion; (**b**) western and southern ecoregions.

The feasibility of the OK to improve Vieilledent's AGB map was assessed by analyzing the evolution in the accuracy of Vieilledent's AGB map after the integration of GLAS data. This evolution was determined by comparing the precision of Vieilledent's AGB map to that of the improved AGB map (Vieilledent's AGB map with the integration of GLAS data). As for the improved AGB map, the precision of Vieilledent's AGB map was calculated using AGB pixels that cover at least two in situ AGB values. The precision of Vieilledent's AGB map and the improved AGB map were calculated using 128 samples (each sample was obtained by averaging at least two in situ AGB values). These samples are located in all ecoregions and have values within the range of in situ AGB used to build the models relating in situ AGB to both GLAS and DEM metrics (Figure 6). Figure 6a shows the AGB from Vieilledent's AGB map as a function of the average in situ AGB. Similarly, the comparison between the AGB from the improved AGB map and the average in situ AGB is shown in Figure 6b. The results show that the OK decreased the RMSE of Vieilledent's AGB map by 6.9 t/ha; the RMSE of the improved AGB map is 74.1 t/ha ($R^2$ = 0.71, RRMSE = 28.2%) compared to an RMSE of 81.0 t/ha ($R^2$ = 0.62, RRMSE = 30.8%) for AGB estimates from Vieilledent's AGB map.

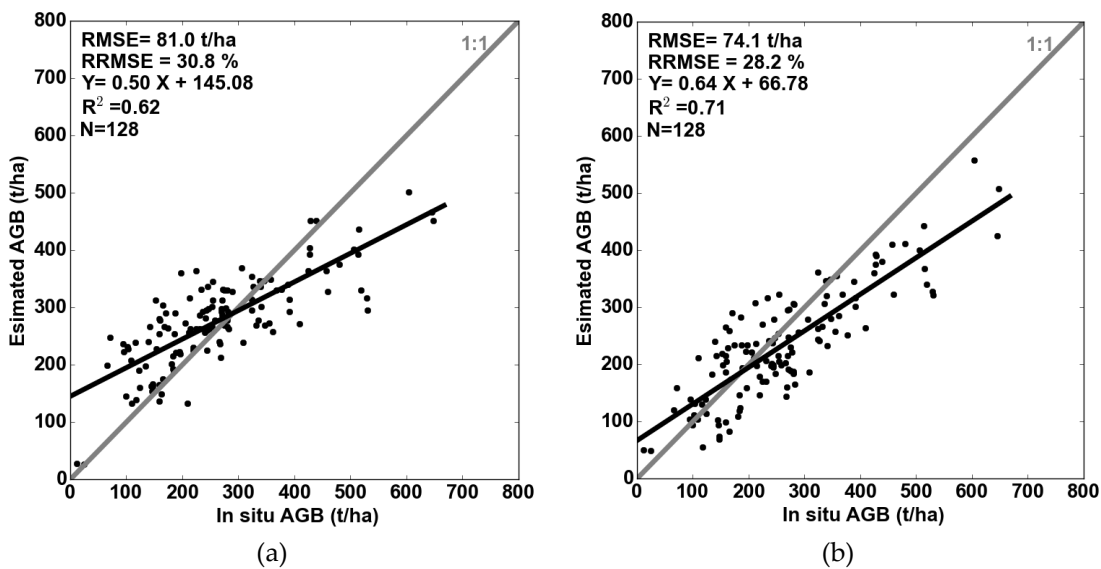

**Figure 6.** (**a**) Vieilledent's AGB map as a function in situ AGB; (**b**) the improved AGB map as a function of in situ AGB. N is the number of samples used for validation.

In addition, statistics were calculated separately for in situ AGB lower and higher than 400 t/ha. For in situ AGB lower than 400 t/ha, the precision of the improved AGB map was better than the precision of Vieilledent's AGB map; the RMSE of the improved AGB map was 61.7 t/ha ($R^2$ = 0.52) and the RMSE of Vieilledent's AGB map was 69.3 t/ha ($R^2$ = 0.49). Similarly, for AGB higher than 400 t/ha, the precision of the improved AGB map (RMSE = 125.6, $R^2$ = 0.34) was better than the precision of Vieilledent's AGB map (RMSE = 131.4 t/ha, $R^2$ = 0.17). Thus, the precision of Vieilledent's AGB map was improved for both lower and higher values of AGB.

*4.3. Comparison between Vieilledent's AGB Map and the Improved AGB Map*

A comparison between Vieilledent's AGB map and the improved AGB map was also performed. At the scale of Madagascar, the results showed that the mean AGB values from both maps are similar (204.8 t/ha for Vieilledent's AGB map, and 197.0 t/ha for the improved AGB map). However, important differences between these maps were observed when the two maps were compared visually (Figure 7). In particular, in the north of the eastern ecoregion, the improved AGB map was able to provides much higher AGB values (up to 647.7 t/ha) compared to Vieilledent's AGB map (maximum AGB pixel value of 529.5 t/ha). In addition, the absolute difference between both maps (the improved AGB map—Vieilledent's AGB map) was more important in the eastern ecoregion than in the western and southern ecoregions (Figure 7c). In the eastern ecoregion, this difference (the improved AGB map—Vieilledent's AGB map) frequently reaches values lower than −100 t/ha and higher than 100 t/ha. For both the western and southern ecoregions, intermediate difference values were obtained (between –40 and 100 t/ha). In addition, the relative difference between the improved AGB map and Vieilledent's AGB map was performed. For the eastern ecoregion, the results showed that the AGB estimates from Vieilledent's AGB map were generally decreased and increased by 25% and 50% at the most, respectively. The increase (up to 50%) was observed mainly for a dense forest stand located in the north of the eastern ecoregion, whereas the decrease (up to 25%) was observed throughout the eastern ecoregion. For both the western and southern ecoregions, the decreases and the increases of the AGB estimates from Vieilledent's AGB map were up to 50% and 100%, respectively.

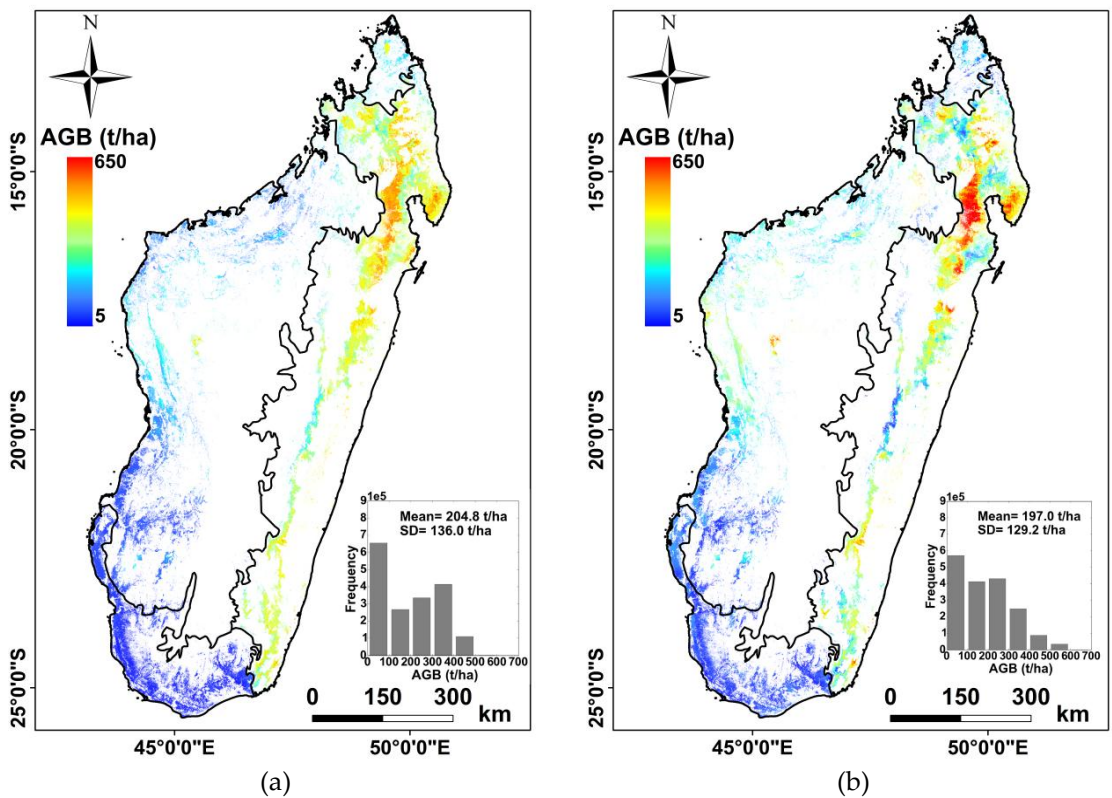

(a) (b)

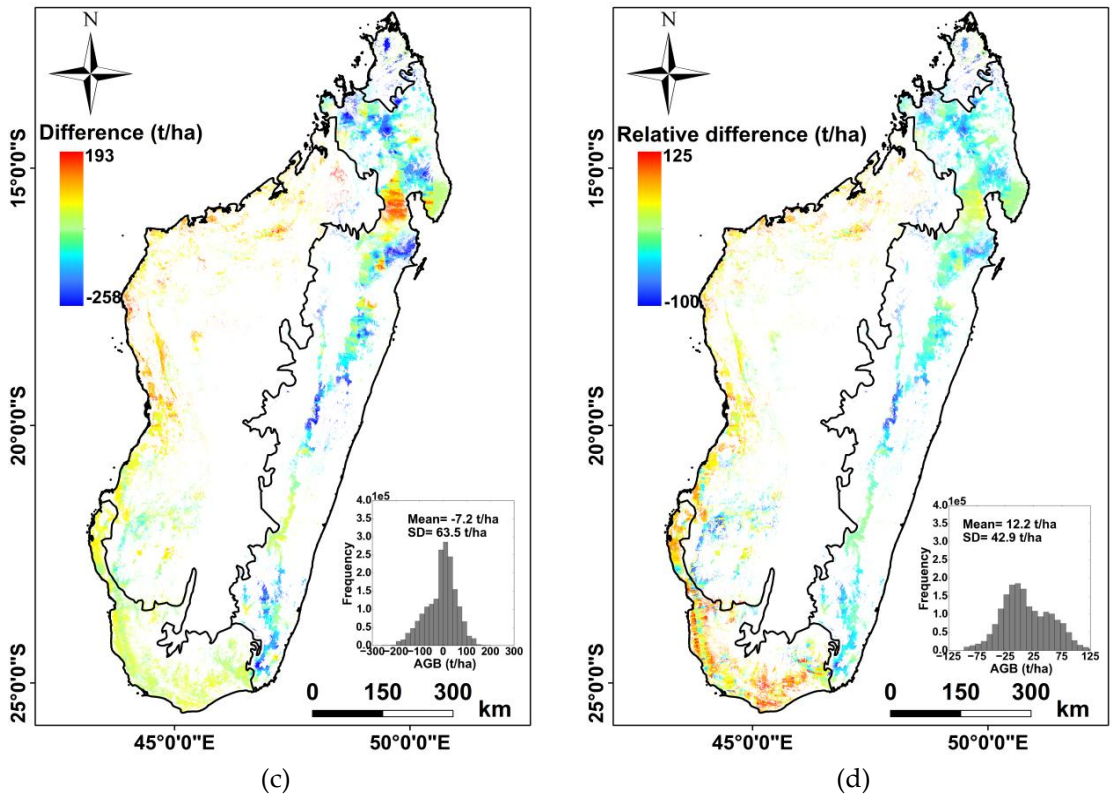

**Figure 7.** (**a**) Vieilledent's AGB map [28] (**b**) the improved AGB map; (**c**) map of the absolute difference (the improved AGB map—Vieilledent's AGB map); (**d**) map of the relative difference. SD: Standard Deviation.

Finally, the carbon stock was computed from both Vieilledent's AGB map and the improved AGB map. To compute the carbon stock, the sum of AGB pixel values was first multiplied by 0.47 to convert AGB to carbon stock, and then multiplied by 6.25 to convert from hectares to an area of 250 m × 250 m (the improved AGB map pixel resolution). The carbon stock in Madagascar from the improved AGB map was estimated to be $0.85964 \times 10^{15}$ PgC (1 PgC = $10^{15}$ grams carbon) compared to $0.89350 \times 10^{15}$ from Vieilledent's AGB map.

*4.4. Comparison between the Improved AGB Map and Avitabile's AGB Map*

Avitabile's AGB map and the improved AGB map resampled to a spatial resolution of 1 km were compared to in situ AGB values (Figure 8). The samples used to build the models for AGB estimation from GLAS and DEM metrics were not used to validate the improved AGB with a spatial resolution of 1 km. The results showed that Avitabile's AGB map is much less accurate (RMSE = 168.9 t/ha, RRMSE = 53.9%, $R^2$ = 0.04) than the improved AGB map with a 1 km spatial resolution (RMSE = 68.9 t/ha, RRMSE = 22.0%, $R^2$ = 0.80).

Finally, the carbon stock was calculated from Avitabile's AGB map and the improved AGB map with a spatial resolution of 1 km. The carbon stock from Avitabile AGB was 1.08548 PgC, which is considerably higher than the carbon stock value calculated from the improved AGB map with spatial resolution of 1 km (0.85628 PgC).

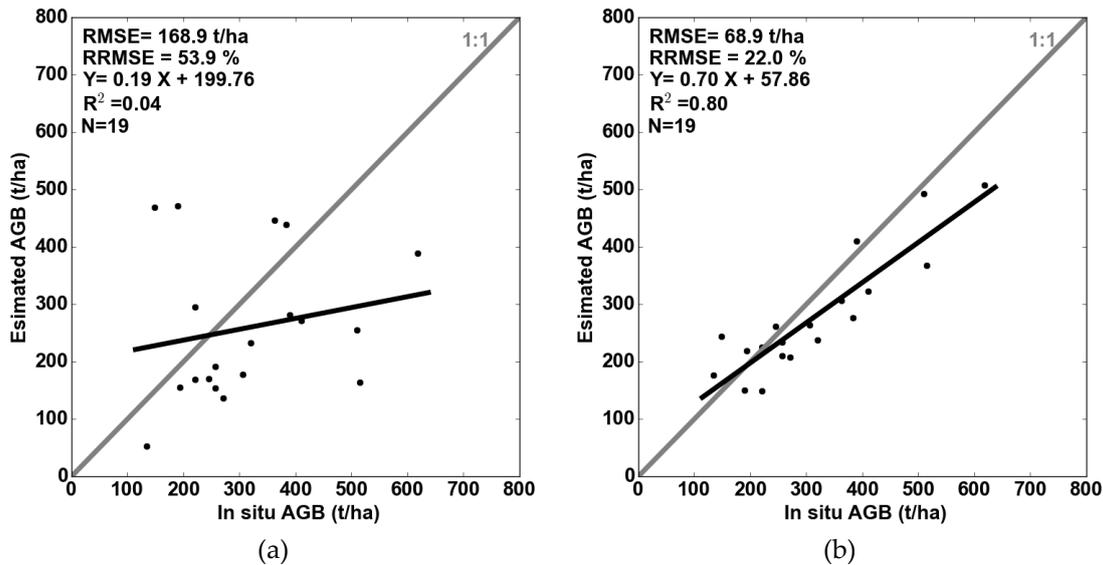

**Figure 8.** (**a**) Avitabile's AGB map as a function of in situ AGB; (**b**) the improved AGB map with 1 km resolution as a function of in situ AGB. N is the number of samples used for validation.

*4.5. Comparison between the Improved AGB Map and the Baccini's Approach AGB Map*

In this section, the precision of the improved AGB map and the precision of the AGB map from Baccini's approach were compared. The results show that the AGB map from Baccini's approach has quite low precision, with a RMSE of 135 t/ha (RRMSE = 51.5%), in comparison to the precision of the improved AGB map, with a RMSE of 74.1 t/ha (RRMSE = 28.2%) (Figure 9). For the AGB map from Baccini's approach, the RMSE is equal to 118.1 t/ha and 212.2 t/ha for in situ AGB lower and higher than 400 t/ha, respectively. In addition, the AGB map from Baccini's approach was not able to map AGB higher than 500 t/ha (value obtained from the AGB map using Baccini's approach).

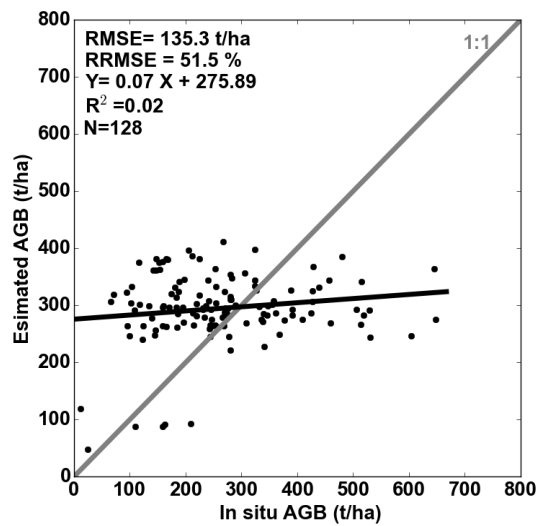

**Figure 9**. Baccini's approach AGB map as a function of in situ AGB.

## 5. Discussion

In this study, a method based on the use of GLAS data was developed mainly to overcome the saturation at high AGB values of the existing AGB map derived by Vieilledent et al. [28] using optical images, parameters computed from DEM, climatic variables, and field inventories. A good correlation between in situ AGB and GLAS and DEM metrics was obtained ($R^2 = 0.91$ for the eastern ecoregion, and $R^2 = 0.66$ for the western and southern ecoregions). Despite the fact that the relationships between in situ AGB and both GLAS and DEM metrics was built with only a few in situ

AGB samples (14 for the eastern ecoregion, and 13 for the western and southern ecoregions) located within a distance of 250 m from GLAS footprints, the integration of GLAS data leads to an improvement in the precision of Vieilledent's AGB map (Figure 6). This improvement was possible because Vieilledent's AGB map was created by using the EVI (Enhanced Vegetation Index) computed from optical images as an input parameter. The EVI saturates at higher AGB levels, leading to an underestimation of high AGB values, which limits the estimation of AGB to lower values (lower than 550 t/ha in the study of Vieilledent et al. [28]). In contrast, our approach is based on the integration of LiDAR GLAS data, which are able to provide much higher AGB values, and thus improve Vieilledent's AGB map (derived using optical data), even in densely forested areas (*in situ* AGB up to 650 t/ha). However, results showed that the improved AGB map values still underestimate the in situ AGB values for AGB higher than 350 t/ha. This is probably due to the fact that (1) the used spaceborne LiDAR data are not sufficiently dense (0.5 points/km$^2$) to completely eliminate the high underestimation of AGB for values higher than 350 t/ha and (2) the data used, mainly optical (such as EVI), to derive the Vieillendent's AGB map have low sensitivity to high AGB values (high than 350 t/ha). Fayad et al. [42] showed that the use of dense spaceborne and airborne LiDAR data in addition to other data (optical, DEM, and environmental) provides good estimation of the AGB. In general, the integration of GLAS-derived AGB moderately increases the precision of Vieilledent's AGB map by 7.6 t/ha and 5.8 t/ha for AGB values lower and higher than 400 t/ha, respectively. Considering in our method remote sensing variables that describe the floristic composition and the biogeography characteristics of forested areas would improve the precision of GLAS-derived AGB and consequently, the correction factor map, and thus brought a larger improvement to Vieilledent's AGB map. Finally, our approach could be applied to improve the most recent and accurate pan-tropical AGB map produced by Avitabile et al. [39] because GLAS LiDAR shots cover the pan-tropical forested areas [43]. Applying our approach would improve Avitabile's AGB map mainly by overcoming the problem of saturation at high AGB values.

Our approach uses an existing AGB map (Vieilledent's AGB map). The use of an existing AGB map does not represent an inconvenience, since it would have been possible for us to derive this AGB map from freely available remote sensing data. The advantage of the proposed approach is that despite LiDAR data density of 0.5 points/km$^2$, a simple ordinary kriging seems sufficient to improve an AGB map derived using optical data (such as Vieilledent's AGB map). It should be noted that GLAS footprints have a good distribution over forested areas in Madagascar. The main potential limitation of our approach is the unavailability of GLAS data since 2009 (GLAS data are between 2003 and 2009). Thus, this method could be applied to improve an AGB map created using data from between 2003 and 2009 and to an AGB map for forests with conditions that have not changed with respect to the period between 2003 and 2009. In addition, field inventories are necessary to apply our approach. However, a relatively low number of field inventories could be enough to calibrate LiDAR and SRTM metrics data to AGB [25].

In this paper, we also compared the improved AGB map and the most recent pan-tropical AGB map provided by Avitabile et al. [39]. Avitabile's AGB map was created using a fusion of data from Baccini's and Saatchi's AGB maps and reference data. The results showed that the improved AGB map has better precision than Avitabile's AGB map. This is because the reference data used for the calibration of the fusion model in the study of Avitabile et al. [39] does not cover the ranges of AGB values of forested areas in Madagascar (up to 700 t/ha according to our in situ AGB measurements). These reference data used by Avitabile et al. [39] (60 samples) are located in a zone to the north of Madagascar characterized by in situ AGB values lower than 235 t/ha [39].

In addition, the most popular method, proposed by Baccini et al. [26], for AGB mapping was applied using our in situ AGB measurements to produce an AGB map (Baccini's approach AGB map). The results showed that the accuracy of the Baccini's approach AGB map was lower than the accuracy of the improved AGB map. Thus, it seems that using optical data as predictive variables to derive AGB estimates do not allow to the accurate estimation of high AGB values, even if GLAS-derived AGB values are used to establish the relationship with the optical data.

Finally, our approach could be applied to improve the existing global pan-tropical AGB map, mainly by overcoming the problem of saturation at high AGB values, from which most of these recent pan-tropical maps, such as Avitabile's AGB map and Baccini's AGB map, suffer.

## 6. Conclusions

This study analyzed the potential of LiDAR sensor data and ICESat/GLAS data to improve a AGB map recently established in Madagascar (Vieilledent's AGB map) using optical and digital elevation model spatial imagery and climatic variables. First, GLAS data were used to provide AGB estimates at 48,247 footprint locations covering the forested areas in Madagascar between 2003 and 2009 with a density of 0.5 points/km$^2$ of forest (Figure 1). Second, the additional correction factors, which are the difference between Vieilledent's AGB map and GLAS AGB, were calculated. Third, an ordinary kriging interpolation was performed using these additional correction factors to provide a correction factor map. Finally, Vieilledent's AGB map and the correction factor map were summed to improve Vieilledent's AGB map.

The results showed that the precision of the improved AGB map (RMSE = 74.1 t/ha, $R^2$ = 0.71, N = 128) produced is better than that of the Vieilledent's AGB map (RMSE = 81.0 t/ha, $R^2$ = 0.62, N = 128). In addition, results showed that the improved AGB map allows higher estimates of AGB than Vieilledent's AGB map. Indeed, the AGB values in the improved AGB map reach 650 t/ha, whereas the maximum AGB value in Vieilledent's AGB map is 550 t/ha. For the improved AGB map, the number of AGB pixels (size = 250 m × 250 m) with a value higher than 550 t/ha is 13,241, covering an area of approximately 68,917 ha in the eastern ecoregion, out of which 61,865 ha represent one continuous forest stand located to the north of the eastern ecoregion.

Moreover, the results show that our approach provides more precise AGB estimates in comparison to the approaches proposed by Baccini et al. [26] and Avitabile et al. [39]. This is because the in situ AGB measurements used to calibrate the GLAS data, i.e., derive AGB from GLAS metrics, cover the range of all in situ AGB values. In addition, optical data that saturate at higher AGB values were not used in the procedure, which leads to the improvement in Vieilledent's AGB map.

A limitation of our method is the unavailability of new GLAS data because ICESat ceased operations in 2009. However, this method could be applied to improve existing AGB maps constructed for forests before 2010 because data acquired by ICESat between 2003 and 2009 are free. The Global Ecosystem Dynamics Investigation LiDAR (GEDI) mission (launch date in 2019) will ensure LiDAR data with a smaller footprint (25 m) for better mapping of pan-tropical forests.

Finally, we assume that the nominal year of the improved AGB map is 2010 because we used the Vieilledent's AGB map performed for the year of 2010, in addition to spaceborne LiDAR data acquired between 2003 and 2009.

**Acknowledgments:** This research was supported by IRSTEA (National Research Institute of Science and Technology for Environment and Agriculture) and the French Space Study Center (CNES, TOSCA 2016). G.V. was supported by the JRC through the ReCaREDD project funded by the European Commission and by the Cirad through the BioSceneMada project funded by FRB/FFEM (project agreement AAP-SCEN-2013 I). The authors wish to thank the institutes in Madagascar (DGF, ONE, WWF, CI, WCS, Goodplanet/ETC Terra and Cirad) for providing free access to the large AGB dataset derived from various forest inventories in Madagascar. The authors also acknowledge the National Snow and Ice Data Center (NSDIC) for the distribution of the ICESat/GLAS data.

**Author Contributions:** M. El Hajj, N. Baghdadi, G. Vieilledent and I. Fayad conceived and designed the experiments; M. El Hajj performed the experiments; M. El Hajj and N. Baghdadi analyzed the data; I. Fayad, G. Vieilledent, J.-S. Bailly and D.H. Tong Minh revised and improved the manuscript; M. El Hajj wrote the manuscript.

**Conflicts of Interest:** The authors declare no conflict of interest.

*Remote Sens.* **2017**, *9*, 213 19 of 1942. Fayad, I.; Baghdadi, N.; Guitet, S.; Bailly, J.-S.; Hérault, B.; Gond, V.; El Hajj, M.; Minh, D.H.T. Aboveground biomass mapping in French Guiana by combining remote sensing, forest inventories and environmental data. *Int. J. Appl. Earth Obs. Geoinf.* **2016**, *52*, 502–514.
43. Baghdadi, N.N.; El Hajj, M.; Bailly, J.-S.; Fabre, F. Viability statistics of GLAS/ICESat data acquired over tropical forests. *IEEE J. Sel. Top. Appl. Earth Obs. Remote Sens.* **2014**, *7*, 1658–1664.
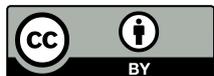

© 2017 by the authors. Licensee MDPI, Basel, Switzerland. This article is an open access article distributed under the terms and conditions of the Creative Commons Attribution (CC-BY) license (http://creativecommons.org/licenses/by/4.0/).